# Ultrasonic imaging with limited-diffraction beams

Jian-yu Lu, Ph.D.
Ultrasound Laboratory, Department of Bioengineering, The University of Toledo, Toledo, OH 43606, U.S.A. E-mail: jilu@eng.utoledo.edu

**Abstract** – Limited-diffraction beams are a class of waves that may be localized in space and time. Theoretically, these beams are propagation invariant and can propagate to an infinite distance without spreading. In practice, when these beams are produced with wave sources of a finite aperture and energy, they have a very large depth of field, meaning that they can keep a small beam width over a large distance. Because of this property, limited-diffraction beams may have applications in various areas such as medical imaging and tissue characterization. In this paper, fundamentals of limited-diffraction beams are reviewed and the studies of these beams are put into a unified theoretical framework. Theory of limited-diffraction beams is further developed. New limited-diffraction solutions to Klein-Gordon Equation and Schrodinger Equation, as well as limited-diffraction solutions to these equations in confined spaces are obtained. The relationship between the transformation that converts any solutions to an ($N$-1)-dimensional wave equation to limited-diffraction solutions of an $N$-dimensional equation and the Lorentz transformation is clarified and extended. The transformation is also applied to the Klein-Gordon Equation. In addition, applications of limited-diffraction beams are summarized.

## I. INTRODUCTION

One type of limited-diffraction beams was first described by Stratton as "undistorted progressive waves" (UPW) in his book in 1941 [1]. In 1987, without referring to Stratton's work, Durnin et al. studied the UPWs with both computer simulation and an optical experiment [2]-[4]. Because the UPWs in Stratton's book have a Bessel transverse beam profile, they are termed Bessel beams. Durnin et al. have named the Bessel beams "nondiffracting beams" or "diffraction-free" beams [2]-[4]. Because Durnin's terminologies are controversial in scientific community, these beams are termed "limited-diffraction beams" [5] since all practical beams or waves will eventually diffract. Bessel beams are localized in transverse direction and may have potential applications [6]-[14]. In acoustics, the first Bessel annular array transducer was designed and constructed in 1990 [15]-[16], and patented in 1992 [17]. The applications of Bessel beams in acoustics were also studied extensively [18]-[30].

Localized waves (LWs) were first developed by Brinttingham in 1983 and were termed focus wave modes [31]. LWs have similar properties as Bessel beams in terms of transverse localization. In addition, LWs contain multiple frequencies and may be localized in the axial direction. LWs have been studied by many investigators [32]-[40]. However, LWs are not propagation invariant, i.e., they do not meet the propagation-invariant condition as defined in the following: If one travels with the wave at a speed, $c_1$, he/she sees a wave packet, $\Phi(\vec{r},t) = \Phi(x, y, z - c_1 t)$, that is unchanged for $z - c_1 t =$constant, where $z$ is the axial axis along the direction of wave propagation, $\vec{r} = (x, y, z)$ is a point in space, and $t$ is the time.

To find multiple frequency waves that are propagation invariant, i.e., $\Phi(\vec{r},t) = \Phi(x, y, z - c_1 t)$, in 1991, X waves were developed [41]-[43] and were subsequently studied [44]-[54]. The name, "X waves", was used because the beam profile in the axial cross-section (a plane through the beam axis) resembles the letter "X". Due to the interest of X waves in nonlinear optics and other applications, X waves were introduced in the "Search and Discovery" column of Physics Today in 2004 [55]. The two 1992 X wave papers [42]-[43] were awarded by the Ultrasonics, Ferroelectrics, and Frequency Control (UFFC) Society of the Institutes of Electrical and Electronics Engineers (IEEE) in 1993. Later, an X wave experiment in optics was performed by Saari and Reivelt and was published in 1997 in Physical

Review Letters [56]. To generalize the X waves, a transformation that is used to obtain limited-diffraction beams (including X waves) in an $N$-dimensional space from any solutions to an ($N$-1)-dimensional isotropic/homogeneous wave equation was developed in 1995 [44], where $N \geq 2$ is an integer. This formula has been related to part of the Lorentz transformation [57]-[58], and was used and demonstrated by other researchers [59]-[60]. Furthermore, an X wave transform that is a transformation pair was developed in 2000 for any physically realizable waves using the orthogonal property of X waves [46]-[47]. The orthogonal property of X waves was further studied by Salo et al. in 2001 [61]. The transformation pair allows one to decompose an arbitrary physically realizable wave into X waves (inverse X wave transform) and determine the coefficients (forward X wave transform) of the decomposition. Based on the X wave theory, a method and its extension that are capable of an ultra high frame rate (HFR) two-dimensional (2D) or three-dimensional (3D) imaging were developed in 1997 [62]-[82]. Due to the importance of this method, it was noted as one of the predictions of the 21st century medical ultrasonics in 2000 [83]. After the introduction of X waves in 1991 [41]-[43], these waves have been studied extensively by many investigators [56], [58]-[60], [84]-[118]. There are also some review papers on X waves and their applications [119]-[126].

In this paper, fundamentals of limited-diffraction beams are reviewed and the studies of these beams are put into a unified theoretical framework. Theory of limited-diffraction beams is further developed. New limited-diffraction solutions to Klein-Gordon Equation and Schrodinger Equation, as well as limited-diffraction solutions to these equations in confined spaces are obtained. The relationship between the transformation that converts any solutions to an ($N$-1)-dimensional wave equation to limited-diffraction solutions of an $N$-dimensional equation and the Lorentz transformation is clarified and extended. The transformation is also applied to the Klein-Gordon Equation. In addition, some applications of limited-diffraction beams are summarized.

## II. FUNDAMENTALS OF LIMITED-DIFFRACTION BEAMS

*A. Bessel Beams*

An $N$-dimensional isotropic/homogeneous wave equation is given by:

$$\left[\sum_{j=1}^{N} \frac{\partial^2}{\partial x_j^2} - \frac{1}{c^2}\frac{\partial^2}{\partial t^2}\right]\Phi(\vec{r},t) = 0, \qquad (1)$$

where $x_j, (j = 1, 2, ..., N)$ represents rectangular coordinates in an $N$-dimensional space, $N \geq 1$ is an integer, $\Phi(\vec{r},t)$ is a scalar function (sound pressure, velocity potential, or Hertz potential in electromagnetics) of spatial variables, $\vec{r} = (x_1, x_2, \cdots x_N)$, and time, $t$, and $c$ is the speed of light in vacuum or speed of sound in a medium.

In three–dimensional space, we have:

$$\left(\nabla^2 - \frac{1}{c^2}\frac{\partial^2}{\partial t^2}\right)\Phi(\vec{r},t) = 0, \qquad (2)$$

where $\nabla^2$ is the Laplace operator. In cylindrical coordinates, the wave equation is given by:

$$\left[\frac{1}{r}\frac{\partial}{\partial r}(r\frac{\partial}{\partial r}) + \frac{1}{r^2}\frac{\partial^2}{\partial \phi^2} + \frac{\partial^2}{\partial z^2} - \frac{1}{c^2}\frac{\partial^2}{\partial t^2}\right]\Phi(\vec{r},t) = 0, \qquad (3)$$

where $r = \sqrt{x^2 + y^2}$ is the radial distance, $\phi = \tan^{-1}(y/x)$ is the polar angle, and $z$ is the axial axis.

One generalized solution to the $N$-dimensional wave equation in (1) is given by [42], [119]:

$$\Phi(x_1, x_2, ..., x_N; t) = f(s), \quad (4)$$

where

$$s = \sum_{j=1}^{N-1} D_j x_j + D_N(x_N \pm c_1 t), \quad (N \geq 1), \quad (5)$$

and where $D_j$ are complex coefficients, $f(s)$ is any well-behaved complex function of $s$, and

$$c_1 = c\sqrt{1 + \sum_{j=1}^{N-1} D_j^2 / D_N^2}. \quad (6)$$

If $c_1$ is real, $f(s)$ and its linear superposition represent limited-diffraction solutions to the $N$-dimensional wave equation (1).

For example, if $N = 3$, $x_1 = x$, $x_2 = y$, $x_3 = z$, $D_1 = \alpha_0(k, \zeta)\cos\theta$, $D_2 = \alpha_0(k, \zeta)\sin\theta$, $D_3 = b(k, \zeta)$, with cylindrical coordinates, one obtains families of solutions to (3) [42], [119]:

$$\Phi_\zeta(s) = \int_0^\infty T(k) \left[ \frac{1}{2\pi} \int_{-\pi}^{\pi} A(\theta) f(s) d\theta \right] dk \quad (7)$$

and

$$\Phi_K(s) = \int_{-\pi}^{\pi} D(\zeta) \left[ \frac{1}{2\pi} \int_{-\pi}^{\pi} A(\theta) f(s) d\theta \right] d\zeta, \quad (8)$$

where

$$s = \alpha_0(k, \zeta) r \cos(\phi - \theta) + b(k, \zeta)[z \pm c_1(k, \zeta) t], \quad (9)$$

and where

$$c_1(k, \zeta) = c\sqrt{1 + [\alpha_0(k, \zeta) / b(k, \zeta)]^2}, \quad (10)$$

and $\alpha_0(k, \zeta)$, $b(k, \zeta)$, $A(\theta)$, $T(k)$, and $D(\zeta)$ are well-behaved functions, and $\theta$, $k$, and $\zeta$ are free parameters. If $c_1(k, \zeta)$ is real, and is not a function of $k$ and $\zeta$, respectively, $\Phi_\zeta(s)$, $\Phi_K(s)$, are families of limited-diffraction solutions to the wave equation (3).

The following function is also a family of limited-diffraction solution to the wave equation [42], [119]:

$$\Phi_L(r, \phi, z - ct) = \Phi_1(r, \phi) \Phi_2(z - ct), \quad (11)$$

where $\Phi_2(z - ct)$ is any well-behaved function of $z - ct$ and $\Phi_1(r, \phi)$ is a solution to the transverse Laplace equation:

$$\left[\frac{1}{r}\frac{\partial}{\partial r}(r\frac{\partial}{\partial r})+\frac{1}{r^2}\frac{\partial^2}{\partial \phi^2}\right]\Phi_1(r,\phi)=0. \tag{12}$$

If $T(k)=\delta(k-k')$, $f(s)=e^s$, $\alpha_0(k,\zeta)=-i\alpha$, and $b(k,\zeta)=i\beta$ in (7) and (9), we have:

$$\Phi_\zeta(s)=\left[\frac{1}{2\pi}\int_{-\pi}^{\pi}A(\theta)e^{-i\alpha r\cos(\phi-\theta)}d\theta\right]e^{i(\beta z-\omega t)}, \tag{13}$$

where $\beta=\sqrt{k'^2-\alpha^2}$ is the propagation parameter, $\delta(k-k')$ is the Dirac-Delta function, and $k'=\omega/c>0$ is the wave number and $\omega$ is the angular frequency. If $A(\theta)=i^n e^{in\theta}$, one obtains an $n$ th-order Bessel beam [2]-[4], [15]-[17]:

$$\begin{aligned}\Phi_{B_n}(\vec{r},t)&=\Phi_{B_n}(r,\phi,z-c_1 t)\\ &=e^{in\phi}J_n(\alpha r)e^{i(\beta z-\omega t)},\ (n=0,1,2,\cdots)\end{aligned} \tag{14}$$

where the subscript "$B_n$" means $n$ th-order Bessel beam, $\alpha$ is a scaling parameter, $J_n(\cdot)$ is the $n$ th-order Bessel function of the first kind, and $c_1=\omega/\beta$ is the phase velocity of the wave. It is clear that Bessel beams are single-frequency waves and are localized in transverse direction. The scaling parameter, $\alpha$, determines the degree of localization. Because of this property, Bessel beams can be applied to medical ultrasonic imaging [15]-[21]. Bessel beams are further studied [22]-[30] along with the studies of acoustic transducers and ultrasound waves [127]-[130].

*B. Nonlinear Bessel Beams*

In medical imaging, nonlinear properties are important to provide additional information of diseased tissues. Harmonics of Bessel beams due to the tissue nonlinearity are useful to obtain higher quality images by combining the localized properties of limited-diffraction beams [22]-[23].

*C. "Frozen Waves"*

It is clear from (14) that, single-frequency Bessel beams have two free parameters. One is the order of the Bessel function and the other is the scaling parameter that changes the phase velocity of the Bessel beams. The order of the Bessel beams, $n$, in (14) has been exploited to produce various limited-diffraction beams of different transverse beam profiles since 1995 [29]-[30]. Another parameter, the scaling parameter, $\alpha$, in (14), has also been used for a linear superposition of Bessel beams to produce a beam of a desired axial profile [24]-[27] for zeroth-order Bessel beams. Although an annular array was used in the production of superposed Bessel beams in these studies, the number of annuli and the width of each ring are free to change. When the number of annuli approaches to infinity and the width of each ring shrinks to zero with a given circular aperture, the field distribution at the surface of the annular array is in fact a continuous function. In a more general way, one could use X wave transform [28], [46]-[47] to produce a wave whose shape would be close to a desired one under conditions such as the least-square criterion [131] by changing both the order of the beams and the scaling parameter.

In 2004, Zamboni-Rached has developed an analytical relationship between the scaling parameter of Bessel beams and the axial beam profile along the beam axis ($r=0$) for the zeroth-order Bessel beams. The resulting linear superposition of Bessel beams of different scaling parameter, $\alpha$, was called "Frozen Waves" [132]. The method was extended to include superposition over both the scaling parameter and the order of the Bessel beams [133] to better control the transverse beam profile of the "frozen waves". These studies not only provide computationally efficient ways for beam designs, but also may have applications in optical tweezers [134].

*D. X Waves*

If $T(k) = B(k)e^{-a_0 k}$, $A(\theta) = i^n e^{in\theta}$, $\alpha_0(k,\zeta) = -ik\sin\zeta$, $b(k,\zeta) = ik\cos\zeta$, and $f(s) = e^s$, one obtains an $n$th-order X wave [41]-[53], which is a superposition of limited-diffraction portion of Axicon beams [135]-[136]:

$$\Phi_{X_n}(\vec{r},t) = \Phi_{X_n}(r,\phi,z-c_1 t)$$
$$= e^{in\phi}\int_0^\infty B(k) J_n(kr\sin\zeta) e^{-k[a_0 - i\cos\zeta(z-c_1 t)]} dk, \quad (n=0,1,2,\cdots) \tag{15}$$

where the subscript "$X_n$" means $n$th-order X wave, $c_1 = c/\cos\zeta \geq c$ is both the phase and group velocity of the wave, $|\zeta| < \pi/2$ is the Axicon angle [136]-[137], $a_0$ is a positive free parameter that determines the decaying speed of the high frequency components of the wave, and $B(k)$ is an arbitrary well-behaved transfer function of a device (acoustic transducer or electromagnetic antenna) that produces the wave. Compare (15) with (14), it is easy to see the similarity and difference between a Bessel beam and X wave. X waves are multiple-frequency waves while Bessel beams have a single frequency. However, both waves have the same limited-diffraction property, i.e., they are propagation invariant. Because of multiple frequencies, X waves can be localized in both transverse space and time to form a tight wave packet. They can propagate in a free space or isotropic/homogeneous media without spreading or dispersion. Choosing specific $B(k)$, one can obtain analytical X wave solutions [41]-[43]. One example is the zeroth-order ($n=0$ and $B(k) = a_0$) X wave [42]:

$$\Phi_{X_0}(\vec{r},t) = \Phi_{X_0}(r,\phi,z-c_1 t)$$
$$= \int_0^\infty a_0 J_0(kr\sin\zeta) e^{-k[a_0 - i\cos\zeta(z-c_1 t)]} dk \tag{16}$$
$$= \frac{a_0}{\sqrt{(r\sin\zeta)^2 + [a_0 - i\cos\zeta(z-c_1 t)]^2}}$$

*E. Obtaining Limited-diffraction Beams with Variable Transformation*

If $\Phi_{N-1}(\vec{r}_{N-1},t)$ is a solution to the ($N-1$)-dimensional isotropic/homogeneous wave equation, one can always obtain a limited-diffraction solution, $\Phi_N(\vec{r}_N, t)$, to the $N$-dimensional wave equation (see (1)) with the following variable substitutions [44]:

$$\begin{cases} \vec{r}_{N-1}\sin\zeta \to \vec{r}_{N-1} \\ \dfrac{x_N \cos\zeta}{c} - t \to t \end{cases} \text{ or } \begin{cases} \vec{r}_{N-1}\sin\zeta \to \vec{r}_{N-1} \\ t - \dfrac{x_N \cos\zeta}{c} \to t \end{cases}, \tag{17}$$

where $\vec{r}_{N-1} = (x_1, x_2, \cdots, x_{N-1})$, $\vec{r}_N = (x_1, x_2, \cdots, x_N)$ and $N \geq 2$ is an integer and $|\zeta| < \pi/2$ is the Axicon angle [136]-[137] (for $N=1$, $\Phi_{N-1}(\vec{r}_{N-1},t) = \Phi_0(t)$ is a vibration and not a wave, in this case, (17) and the procedure above work only when $\zeta = 0$). Because $x_N \cos\zeta/c - t$ appears as a single variable in the following equation:

$$\Phi_N(\vec{r}_N, t) = \Phi_N(\vec{r}_{N-1}, x_N - c_1 t) = \Phi_{N-1}(\vec{r}_{N-1}\sin\zeta, x_N \cos\zeta/c - t), \tag{18}$$

$\Phi_N(\vec{r}_N,t)$ is a limited-diffraction beam propagating along the axis, $x_N$. As shown in [57]-[58], (17) is related to part of the Lorentz transformation (missing the transformation on $x_N$) after dividing all variables by the same constant, $\sin\zeta$:

$$\begin{cases} \vec{r}_{n-1} \to \vec{r}_{n-1} \\ \dfrac{t}{\sin\zeta} - \dfrac{x_n \cos\zeta}{c\sin\zeta} = \dfrac{1}{\sin\zeta}(t - \dfrac{x_n \cos\zeta}{c}) = \gamma(t - \dfrac{\beta}{c}x_n) \to t' \end{cases} \quad (19)$$

where $\beta = \cos\zeta = v/c$ and $\gamma = \dfrac{1}{\sin\zeta} = \dfrac{1}{\sqrt{1-\beta^2}}$, and where $0 \le v < c$ is the velocity of the moving coordinates (observer) along the axis, $x_N$. Contrary to the report in [57]-[58], if $\Phi_{N-1}(\vec{r}_{N-1},t)$ is a solution to the ($N$-1)-dimensional isotropic/homogeneous wave equation, $\Phi_N(\vec{r}_N,t)$ will not be a solution to the $N$-dimensional wave equation (1) with the partial Lorentz transformation (19). Equation (17) has also been used in [59]-[60] to derive limited-diffraction beams in waveguides.

*F. Limited-diffraction Solutions to Klein-Gordon Equation*

An $N$-dimensional Klein-Gordon Equation for a free relativistic particle is given by [138]:

$$\left[\sum_{j=1}^{N} \frac{\partial^2}{\partial x_j^2} - \frac{1}{c^2}\frac{\partial^2}{\partial t^2} - \frac{m^2 c^2}{\hbar^2}\right]\Phi_N(\vec{r}_N,t) = 0, \quad (20)$$

where $x_j, (j=1,2,...,N)$ represents rectangular coordinates in an $N$-dimensional space, $N \ge 1$ is an integer, $\Phi_N(\vec{r}_N,t)$ is a scalar wave function of spatial variables, $\vec{r}_N = (x_1, x_2, \cdots x_N)$, and time, $t$, $c$ is the speed of light in vacuum, $\hbar = h/2\pi$, where $h$ is the Plank constant, $m = m'\sin\zeta$ is the mass of the particle at rest, where $m'$ is a mass related constant and $|\zeta| < \pi/2$ is the Axicon angle [136]-[137].

Assuming $\Phi_{N-1}(\vec{r}_{N-1},t)$ is a solution to the following ($N$-1)-dimensional Klein-Gordon Equation with a mass $m'$ [138]:

$$\left[\sum_{j=1}^{N-1} \frac{\partial^2}{\partial x_j^2} - \frac{1}{c^2}\frac{\partial^2}{\partial t^2} - \frac{m'^2 c^2}{\hbar^2}\right]\Phi_{N-1}(\vec{r}_{N-1},t) = 0, \quad (21)$$

where $\vec{r}_{N-1} = (x_1, x_2, \cdots x_{N-1})$, (18) is a solution to (20) after the variable substitution (17). This can be proved easily in a similar way as that in [44]. Using (18) and (21), we have:

$$\begin{aligned}
&\left[\sum_{j=1}^{N-1} \frac{\partial^2}{\partial x_j^2}\right]\Phi_{N-1}(\vec{r}_{N-1}\sin\zeta, \frac{x_N \cos\zeta}{c} - t) \\
&= \sin^2\zeta \left[\frac{1}{c^2}\frac{\partial^2}{\partial t^2} + \frac{m'^2 c^2}{\hbar^2}\right]\Phi_{N-1}(\vec{r}_{N-1}\sin\zeta, \frac{x_N \cos\zeta}{c} - t)
\end{aligned} \quad (22)$$

and

$$\frac{\partial^2}{\partial x_N^2} \Phi_{N-1}(\vec{r}_{N-1} \sin\zeta, \frac{x_N \cos\zeta}{c} - t)$$
$$= \frac{\cos^2\zeta}{c^2} \frac{\partial^2}{\partial t^2} \Phi_{N-1}(\vec{r}_{N-1} \sin\zeta, \frac{x_N \cos\zeta}{c} - t) \qquad (23)$$

Summing both the left- and right-hand sides of (22) and (23), and comparing the results with (20), it is clear that (18) is a solution to (20). Limited-diffraction solutions to the Klein-Gordon Equation mean that a free relativistic particle may be accompanied by a rigidly propagating wave along the axis, $x_N$, at a velocity that is greater than the speed of light in vacuum in a manner similar to that of X waves [41]-[43] (for $\zeta \neq 0$). If $|\zeta| \to \pi/2$, the wave speed $c_1 = \frac{c}{\cos\zeta} \to \infty$, then one has $m' \to m$. For photons where $m=0$, (22) and (23) are the same as those in [44]. It is worth noting that from the proofs in (22)-(23) and in [44], it is clear that the functions, $\sin\zeta$ and $\cos\zeta$, in (17) can be other functions as long as the summation of the squares of those functions is equal to one: $f_1^2(\zeta) + f_2^2(\zeta) \equiv 1$, where $f_1(\zeta)$ and $f_2(\zeta)$ are any well-behaved functions of $\zeta$ or other free parameters. This extends the transformation formula in (17).

In the following, we will obtain some localized limited-diffraction solutions to the Klein-Gordon Equation. Assuming $f(s) = e^s$ in (4), where $s$ is given by (5), and inserting (4) into (20), one obtains the velocity of the wave:

$$c_1 = c \sqrt{\left(\sum_{j=1}^{N} D_j^2 - \frac{m^2 c^2}{\hbar^2}\right) / D_N^2} . \qquad (24)$$

If $N=3$, $x_1 = x$, $x_2 = y$, and $x_3 = z$, (24) becomes:

$$c_1 = c \sqrt{\left(D_1^2 + D_2^2 + D_3^2 - \frac{m^2 c^2}{\hbar^2}\right) / D_3^2} . \qquad (25)$$

Choosing $D_1 = \alpha_0 \cos\theta$ and $D_2 = \alpha_0 \sin\theta$, where $-\pi \leq \theta \leq \pi$ is a free parameter and $\alpha_0$ is a well-behaved function of any free parameters, if $\alpha_0 = -i\frac{mc}{\hbar}\sin\zeta$, one obtains:

$$D_3 = i\frac{mc}{\hbar}\left(\sqrt{1 - \left(\frac{\hbar}{mc}\right)^2 \alpha_0^2} \Big/ \sqrt{(c_1/c)^2 - 1}\right) = i\frac{mc}{\hbar}\left(\sqrt{1 + \sin^2\zeta} \Big/ \sqrt{(c_1/c)^2 - 1}\right). \qquad (26)$$

Since $e^s$ in (4) is a solution to the Klein-Gordon Equation (20), a linear superposition over the free parameter, $\theta$, is still a solution:

$$\Phi_{B_n}^{KG}(\vec{r}, t) = \frac{1}{2\pi} \int_{-\pi}^{\pi} A(\theta) e^s d\theta$$
$$= e^{in\phi} J_n\left(\frac{mc}{\hbar} r \sin\zeta\right) e^{i\frac{mc}{\hbar}\left(\sqrt{1+\sin^2\zeta}/\sqrt{(c_1/c)^2-1}\right)(z-c_1 t)}, \quad (n=0,1,2,\cdots) \qquad (27)$$

where the subscript "$B_n$" and the superscript "$KG$" mean $n$th-order Bessel beam and the Klein-Gordon Equation, respectively, and $i\frac{mc}{\hbar}\left(\sqrt{1+\sin^2\zeta}/\sqrt{(c_1/c)^2-1}\right)$ is the propagation constant. Equation (27) is a localized solution to (20) and its localization increases with the mass, $m$. For electron at rest, $m = 9.1 \times 10^{-31}$ kg, and thus $mc/\hbar = 2.6 \times 10^{12}$ m$^{-1}$ ($\hbar = 1.05 \times 10^{-34}$ J·s and $c = 3.0 \times 10^8$ m/s). The wave in (27) is localized in pico-meter scale if $\sin\zeta \approx 1$. There are other choices of $\alpha_0$. If $\alpha_0$ is a constant, a localized limited-diffraction solution that has a fixed transverse beam profile can be obtained. If $\alpha_0 = -i(\gamma mv/\hbar)\sin\zeta$, where $\gamma = 1/\sqrt{1-\beta^2}$ and $\beta = v/c$ and where $v$ is the velocity of the particle, the transverse localization of the solutions will increase with the speed of the particle. In this case, the propagation constant is given by: $i\frac{mc}{\hbar}\left(\sqrt{1+(\gamma v/c)^2 \sin^2\zeta}/\sqrt{(c_1/c)^2-1}\right)$.

Superposing $\Phi_{B_n}^{KG}(\vec{r},t)$ in (27) over the mass, $m$, one obtains a composed wave function that is similar to the X wave [41]-[53], but may not necessary be a solution to (20) where $m$ is a constant for a given particle (the physical meaning could be a group of independent particles of different masses traveling in space). Using (7) and (27), and letting $T(k) = B(k)e^{-a_0 k}$, where $k = mc/\hbar$, one obtains:

$$\Phi_{X_n}^{KG}(\vec{r},t) = \Phi_{X_n}^{KG}(r,\phi,z-c_1 t)$$
$$= \frac{c}{\hbar} e^{in\phi} \int_0^\infty B(\frac{mc}{\hbar}) J_n(\frac{mc}{\hbar} r \sin\zeta) e^{-\frac{mc}{\hbar}\left[a_0 - i\left(\sqrt{1+\sin^2\zeta}/\sqrt{(c_1/c)^2-1}\right)(z-c_1 t)\right]} dm, (28)$$
$$(n = 0, 1, 2, \cdots)$$

where the subscript "$X_n$" means $n$th-order X wave, the superscript "$KG$" represents Klein-Gordon Equation, $a_0$ is a positive free parameter, and $B(k)$ is an arbitrary well-behaved transfer function. If $n=0$ and $B(k) = a_0$, from (28) and (16) one has (where $c_1$ is a constant) [42]:

$$\Phi_{X_0}^{KG}(\vec{r},t) = \Phi_{X_0}^{KG}(r,\phi,z-c_1 t)$$
$$= \frac{a_0}{\sqrt{(r\sin\zeta)^2 + \left[a_0 - i\left(\sqrt{1+\sin^2\zeta}/\sqrt{(c_1/c)^2-1}\right)(z-c_1 t)\right]^2}}. (29)$$

It is clear from (26)-(29) that, if $c_1 < c$, the solutions or functions are no longer waves. If $c_1 = c$, $D_3$ in (26) is infinity. For $c_1 > c$, one obtains rigidly propagating superluminal waves or functions as in the case of X waves [42]. One example is to assume $c_1 = c/\cos\zeta$ as given in (17) [44]. A superposition that is similar to (28) can also be done over the velocity, $v$, instead of the mass, $m$, of a particle if, say, $\alpha_0 = -i(\gamma mv/\hbar)\sin\zeta$. In this case, the superposition is a limited-diffraction solution to the Klein-Gordon Equation (20).

*G. Limited-diffraction Solutions to Schrodinger Equation*

The general nonrelativistic, time-dependent, and three-dimensional Schrodinger wave equation for multiple particles is given by (see e.g., [139]):

$$-\sum_{j=1}^{M} \frac{\hbar^2}{2m_j} \nabla_j^2 \Phi + V\Phi = i\hbar \frac{\partial \Phi}{\partial t}, \tag{30}$$

where $\Phi = \Phi(x_1, x_2, x_3; \ldots; x_{3M-2}, x_{3M-1}, x_{3M}; t)$ is the wave function (related to the probability of finding particles in space and time) and $V = V(x_1, x_2, x_3; \ldots; x_{3M-2}, x_{3M-1}, x_{3M}; t)$ is the potential of the system. $\Phi$ and $V$ are determined by all the particles and their interactions. $\nabla_j^2$ is the Laplace in terms of the position of the $j$th particle in space, $\vec{r}_j = (x_{3j-2}, x_{3j-1}, x_{3j})$, where $j = 1, 2, \cdots, M$, $M$ is an integer. $m_j$ is the mass at rest of the $j$th particle. Assuming that $V$ is not a function of spatial variables and time, and $\Phi(s) = e^s$, where $s$ is given by (5), one obtains [54]:

$$c_1 = \frac{\sum_{j=1}^{M} \frac{-\hbar^2}{2m_j}\left[D_{3j-2}^2 + D_{3j-1}^2 + D_{3j}^2\right] + V}{-i\hbar D_{3M}}. \tag{31}$$

If $M = 1$, $x_1 = x$, $x_2 = y$, $x_3 = z$, $m_1 = m$, $D_1 = \alpha_0 \cos\theta$, and $D_2 = \alpha_0 \sin\theta$, where $|\zeta| < \pi/2$ is an Axicon angle, $-\pi \leq \theta \leq \pi$ is a free parameter, and $\alpha_0$ is a well-behaved function of any free parameters, (31) is simplified [54]:

$$c_1 = \frac{\frac{-\hbar^2}{2m}\left(\alpha_0^2 + D_3^2\right) + V}{-i\hbar D_3}. \tag{32}$$

If $V = 0$ and $\alpha_0 = -i\frac{mc}{\hbar}\sin\zeta$, one has:

$$D_3 = \begin{cases} i\frac{mc}{\hbar}\left(c_1/c \pm \sqrt{(c_1/c)^2 + \frac{\hbar^2}{m^2 c^2}\alpha_0^2}\right) = i\frac{mc}{\hbar}\left(c_1/c \pm \sqrt{(c_1/c)^2 - \sin^2\zeta}\right), & \alpha_0 \neq 0 \\ i2\frac{mc}{\hbar}(c_1/c), & \alpha_0 = 0 \end{cases}. \tag{33}$$

Following the steps to obtain (27), one obtains a localized solution to the Schrodinger Equation in (30) under the conditions leading to (33) [54]:

$$\Phi_{B_n}^S(\vec{r}, t) = \frac{1}{2\pi} \int_{-\pi}^{\pi} A(\theta) e^s d\theta$$

$$= e^{in\phi} J_n\left(\frac{mc}{\hbar} r \sin\zeta\right) e^{i\frac{mc}{\hbar}\left(c_1/c \pm \sqrt{(c_1/c)^2 - \sin^2\zeta}\right)(z - c_1 t)}, \quad (n = 0, 1, 2, \cdots) \tag{34}$$

where the subscript "$B_n$" and the superscript "$S$" mean $n$th-order Bessel beam and the Schrodinger Equation, respectively, and $i\frac{mc}{\hbar}\left(c_1/c \pm \sqrt{(c_1/c)^2 - \sin^2\zeta}\right)$ is the propagation constant. Similar to the Klein-Gordon Equation (see the text below (27)), one can select $\alpha_0 = $constant, $\alpha_0 = -i(\gamma m v/\hbar)\sin\zeta$, or other functions to obtain more limited-diffraction beams (the corresponding $D_3$ can be easily obtained by inserting different $\alpha_0$ into (33)).

Following the derivations of (28) and substituting $\frac{mc}{\hbar}\left(\sqrt{1+\sin^2\zeta}/\sqrt{(c_1/c)^2-1}\right)$ with $\frac{mc}{\hbar}\left(c_1/c \pm \sqrt{(c_1/c)^2 - \sin^2\zeta}\right)$, one obtains a function that is similar to the X wave [41]-[53], but may not necessary be a solution to (30) (the physical meaning could be a group of independent particles of different masses traveling in space):

$$\Phi^S_{X_n}(\vec{r},t) = \Phi^S_{X_n}(r,\phi,z-c_1 t)$$

$$= \frac{c}{\hbar} e^{in\phi} \int_0^\infty B(\frac{mc}{\hbar}) J_n(\frac{mc}{\hbar} r \sin\zeta) e^{-\frac{mc}{\hbar}\left[a_0 - i\left(c_1/c \pm \sqrt{(c_1/c)^2 - \sin^2\zeta}\right)(z-c_1 t)\right]} dm \,, \quad (35)$$

$$(n = 0, 1, 2, \cdots)$$

where the subscript "$X_n$" means $n$ th-order X wave, the superscript "$S$" represents Schrodinger Equation, $a_0$ is a positive free parameter, and $B(k)$ is an arbitrary well-behaved transfer function. If $n=0$ and $B(k)=a_0$, from (35) and (16) one obtains (where $c_1$ is a constant) [42]:

$$\Phi^S_{X_0}(\vec{r},t) = \Phi^S_{X_0}(r,\phi,z-c_1 t)$$

$$= \frac{a_0}{\sqrt{(r\sin\zeta)^2 + \left[a_0 - i\left(c_1/c \pm \sqrt{(c_1/c)^2 - \sin^2\zeta}\right)(z-c_1 t)\right]^2}} . \quad (36)$$

In (33)-(36), if $(c_1/c)^2 - \sin^2\zeta < 0$, the solutions or functions are unbounded for some $z$ or $t$ and may not be of interest. If $(c_1/c)^2 - \sin^2\zeta \geq 0$, one obtains limited-diffraction solutions or functions [42]. One example is to assume $c_1 = c/\cos\zeta$ as given in (17) [44]. A superposition that is similar to (35) can also be done over the velocity, $v$, instead of the mass, $m$, of a particle if, say, $\alpha_0 = -i(\gamma m v/\hbar)\sin\zeta$. In this case, the superposition is a limited-diffraction solution to the Schrodinger Equation (30).

*H. Electromagnetic X Waves*

The free-space Maxwell's equations are given by [140]:

$$\begin{cases} \nabla \times \vec{E} = -\mu_0 \frac{\partial \vec{H}}{\partial t} \\ \nabla \times \vec{H} = \varepsilon_0 \frac{\partial \vec{E}}{\partial t} \\ \nabla \cdot \vec{E} = 0 \\ \nabla \cdot \vec{H} = 0 \end{cases} , \quad (37)$$

where $\vec{E}$ is the electric field strength, $\vec{H}$ is the magnetic field strength, $\varepsilon_0$ is the dielectric constant of free space ($\varepsilon_0 \approx \frac{\pi}{36} \times 10^{-9}$ F/m), $\mu_0$ is the magnetic permeability of free space ($\mu_0 = 4\pi \times 10^{-7}$ H/m), and $t$ is the time.

Because of the 3$^{rd}$ equation of (37), the electric field strength can be written [54], [141]:

$$\vec{E} = -\mu_0 \frac{\partial}{\partial t} \nabla \times \vec{\Pi}_m , \quad (38)$$

where $\vec{\Pi}_m = \Phi \vec{n}^0$ is a magnetic Hertz vector potential with transverse electrical (TE) polarization, where $\Phi$ is a scalar function, and $\vec{n}^0$ represents a unit vector. Inserting (38) into the 1$^{st}$ equation of (37), one obtains:

$$\vec{H} = \nabla \times (\nabla \times \vec{\Pi}_m). \tag{39}$$

From (37) to (39), one obtains the following vector wave equation:

$$\nabla^2 \vec{\Pi}_m - \frac{1}{c^2} \frac{\partial^2 \vec{\Pi}_m}{\partial t^2} = 0. \tag{40}$$

Let $\vec{n}^\circ = \vec{z}^\circ$, where $\vec{z}^\circ$ is a unit vector along the $z$-axis and use the cylindrical coordinates, from (38) and (39) one obtains:

$$\vec{E} = -\mu_0 \frac{1}{r} \frac{\partial^2 \Phi}{\partial t \partial \phi} \vec{r}^0 + \mu_0 \frac{\partial^2 \Phi}{\partial t \partial r} \vec{\phi}^0 \tag{41}$$

and

$$\vec{H} = \frac{\partial^2 \Phi}{\partial r \partial z} \vec{r}^0 + \frac{1}{r} \frac{\partial^2 \Phi}{\partial \phi \partial z} \vec{\phi}^0 + \left( \frac{\partial^2 \Phi}{\partial z^2} - \frac{1}{c^2} \frac{\partial^2 \Phi}{\partial t^2} \right) \vec{z}^0, \tag{42}$$

respectively, where $\Phi$ is a solution to the free-space scalar wave equation (2), and where $\vec{r}^0$ and $\vec{\phi}^0$ are the unit vectors along the variables, $r$ and $\phi$, respectively. Once a solution to (2) is found, the electrical field strength, $\vec{E}$, and the magnetic field strength, $\vec{H}$, can be obtained from Eqs. (41) and (42), respectively.

If $\Phi$ is an $n$th-order broadband X wave solution or a general X wave solution (see (15)) to (2), the components of $\vec{E}$ and $\vec{H}$ are also X wave type of functions [54]. From the $\vec{E}$ and $\vec{H}$ expressions, the Poynting energy flux vector and the energy density can be derived [54]. The solution of $\vec{E}$ and $\vec{H}$ obtained this way will be limited-diffraction solutions to the Maxwell's equations in (37) [54].

*I. Limited-diffraction Beams in Confined Spaces*

Limited-diffraction beams in confined spaces are of interest [59]-[60], [142]. Previously, Shaarawi et al. [143] and Ziolkowski et al. [144] have shown that the "localized waves" such as Focused Wave Modes and Modified Power Spectrum Pulses, etc., can also propagate in waveguides for an extended propagation depth. In the following, theoretical results of X waves propagating in a confined space such as a waveguide will be developed for acoustics, electromagnetics, and quantum mechanics [142].

*1. Acoustic Waves:* Assuming that $\Phi$ in (2) represents acoustic pressure in an infinitely long cylindrical acoustical waveguide (radius $a$), which is filled with an isotropic/homogeneous lossless fluid medium enclosed in an infinitely rigid boundary. In this case, the normal vibration velocity of the medium at the wall of the cylindrical waveguide is zero for all the frequency components of the X waves, i.e., $\frac{\partial}{\partial r} \Phi_{X_n}(\vec{r}, t; \omega) \equiv 0$, $\forall \omega \geq 0$ at $r = 0$, where $\Phi_{X_n}(\vec{r}, t; \omega)$ is the X wave component at the angular frequency, $\omega$ (see (15)). To meet this boundary condition, the parameter, $k$, in (15) is quantized:

$$k_{nj} = \frac{\mu_{nj}}{a \sin \zeta}, \quad (n, j = 0, 1, 2, \cdots), \tag{43}$$

where $\mu_{nj}$ are the roots of the equations:

$$\begin{cases} J_1(x) = 0, & n = 0 \\ J_{n-1}(x) = J_{n+1}(x), & n = 1, 2, \cdots \end{cases} \quad (44)$$

Thus, the integral in (15) can be changed to a series representing frequency quantized X waves [142]:

$$\Phi_{X_n}(\vec{r},t) = e^{in\phi} \sum_{j=0}^{\infty} \Delta k_{nj} B(k_{nj}) J_n(k_{nj} r \sin \zeta) e^{-k_{nj}[a_0 - i\cos\zeta(z-c_1 t)]}, \quad r \leq a, \quad (45)$$

$$(n = 0, 1, 2, \cdots)$$

where $\Delta k_{n0} = k_{n1}$ and $\Delta k_{nj} = k_{nj+1} - k_{nj}$ ($j = 1, 2, 3, \cdots$). Unlike conventional guided waves, the frequency quantized X waves contain multiple frequencies and propagate through waveguides at the speed of $c_1$ without dispersion. It is noticed that similar results can also be obtained for waveguides of other homogeneous boundary conditions. For an infinitely long cylindrical acoustical waveguide consisting of isotropic/homogeneous lossless media in a free space (vacuum) with a radius $a$, the acoustical pressure is zero at the boundary of the waveguide, $r = a$, i.e., $\mu_{nj}$, ($n, j = 1, 2, 3, \cdots$), in (43) are roots of $J_n(x) = 0$, ($j = 1, 2, \cdots$). See Figs. 1 to 3 for an example of X waves in an acoustic waveguide [142].

It is clear that if $n = 0$, (45) represents an axially symmetric frequency quantized X wave. If $a \to \infty$, then $\Delta k_{nj} \to 0$ and the summation in (45) becomes an integration that represents the X waves in (15). On the other hand, if $a \to 0$, both $k_{nj}$ and $\Delta k_{nj} \to \infty$ ($n, j = 0, 1, 2, \cdots$). This means that for a small waveguide, only high frequency quantized X waves can propagate through it.

*2. Electromagnetic Waves:* The free-space vector wave equations from the free-space Maxwell's equations (37) are given by [145]:

$$\nabla^2 \vec{E} - \frac{1}{c^2} \frac{\partial^2 \vec{E}}{\partial t^2} = 0, \quad (46)$$

and

$$\nabla^2 \vec{H} - \frac{1}{c^2} \frac{\partial^2 \vec{H}}{\partial t^2} = 0. \quad (47)$$

A solution to (46) can be written as:

$$\vec{E}(\vec{r},t) = \vec{E}_\perp(r,\phi) e^{\gamma z - i\omega t}, \quad (48)$$

where $\gamma = i\beta$ is a propagation constant, $\beta = \sqrt{k^2 - k_c^2} > 0$ (for propagation waves), $k = \omega/c$ is the wave number, and $\vec{E}_\perp(r,\phi)$ is a solution of the transverse vector Helmholtz equation:

$$\nabla_\perp \vec{E}_\perp(r,\phi) + k_c^2 \vec{E}_\perp(r,\phi) = 0, \quad (49)$$

where $\nabla_\perp$ is the transverse Laplace operator and $k_c$ is a parameter that is independent of $r$, $\phi$, $z$, and $t$. For transverse magnetic (TM) waves, $\vec{E}_\perp(r,\phi) = E_z(r,\phi) \vec{z}^{\,0}$ and (49) becomes a scalar Helmholtz equation of $E_z(r,\phi)$, where $\vec{z}^{\,0}$ is a unit vector along the $z$-axis.

If $k_c = k \sin \zeta$ where $|\zeta| < \pi/2$ is a constant, after taking into consideration of the exponential term in (48) and integrating the solution of (49) from 0 to $\infty$ over $k$, one obtains an $n$th-order X wave solution (replace the symbol, $\Phi_{X_n}(\vec{r},t)$, in (15) with $E_{z_{nX}}(\vec{r},t)$, where the subscript "X" means X wave). Assuming that electromagnetic X waves travel in vacuum in a totally conductive cylindrical waveguide of a radius, $a$, (i.e., $E_{z_{nX}}(\vec{r},t) \equiv 0$ at $r = a$), similar to the frequency quantization procedure of the acoustic case (45), one obtains [142]:

$$E_{z_{nX}}(\vec{r},t) = e^{in\phi} \sum_{j=0}^{\infty} \Delta k_{nj} B(k_{nj}) J_n(k_{nj} r \sin\zeta) e^{-k_{nj}[a_0 - i\cos\zeta(z - c_1 t)]}, \quad r \leq a, \quad (50)$$

$$(n = 0, 1, 2, ...)$$

where $k_{nj}$ $(n, j = 0,1,2,\cdots)$ are given by (43), and $\mu_{nj}$ $(n, j = 0,1,2,\cdots)$ in (43) are roots of $J_n(x)=0$ $(n = 0,1,2,\cdots)$. Other components of $\vec{E}$ and $\vec{H}$, can be derived from $E_z(\vec{r},t)$ using the free-space Maxwell's equations (37). They will have the same speed, $c_1$, as $E_z$. For transverse electric (TE) waves, results are similar.

*3. DeBroglie Waves:* With a finite transverse spatial extension (such as a free particle passing through a hole of a finite aperture), the function $\Phi_{X_n}^{KG}(\vec{r},t)$ in (28) or $\Phi_{X_n}^{S}(\vec{r},t)$ in (35) would change (spread or diffract) after certain distance behind the hole. However, in the cases such as particles passing through a pipe, $\Phi_{X_n}^{KG}(\vec{r},t)$ and $\Phi_{X_n}^{S}(\vec{r},t)$ need to meet the boundary conditions that they are zero on the wall of the pipe. This gives the following quantized X wave functions corresponding to (28) and (35) respectively [142],

$$\Phi_{X_n}^{KG}(\vec{r},t) = e^{in\phi} \sum_{j=0}^{\infty} \Delta k_{nj} B(k_{nj}) J_n(k_{nj} r \sin\zeta) e^{-k_{nj}\left[a_0 - i\left(\sqrt{1+\sin^2\zeta}/\sqrt{(c_1/c)^2 - 1}\right)(z - c_1 t)\right]}, \quad r \leq a, \quad (51)$$

$$(n = 0, 1, 2, ...)$$

and

$$\Phi_{X_n}^{S}(\vec{r},t) = e^{in\phi} \sum_{j=0}^{\infty} \Delta k_{nj} B(k_{nj}) J_n(k_{nj} r \sin\zeta) e^{-k_{nj}\left[a_0 - i\left(c_1/c \pm \sqrt{(c_1/c)^2 - \sin^2\zeta}\right)(z - c_1 t)\right]}, \quad r \leq a, \quad (52)$$

$$(n = 0, 1, 2, ...)$$

where $k_{nj} = m_{nj} c / \hbar$ $(n, j = 0,1,2,\cdots)$ are given by (43), and $\mu_{nj}$ $(n, j = 0,1,2,\cdots)$ in (43) are roots of $J_n(x)=0$ $(n = 0,1,2,\cdots)$. (51) and (52) represent particles in a confined space with their quantized de Broglie's waves. The quantization may only allow particles of certain mass to pass through the pipe (waveguide). As mentioned in the text below (27) and (34), the free parameter $\alpha_0$ can be chosen differently. If $\alpha_0 = -i(\gamma m v / \hbar)\sin\zeta$, the quantization in (51) and (52) may be modified for summation over the velocity $v$, instead of $m$, of the particles. In this case, only particles with certain velocities are allowed to pass through the pipe or a small nanotube.

There are other implications of the studies above. As we know, light in free space behaves like a wave, but acts as particles (photons) when interacts with materials. Some microscopic structures of materials could be considered as optical waveguides within which the light waves are confined. From the above discussion of the X waves in confined spaces, it is understood that only the light waves that have a higher energy (or frequency) can penetrate these materials or to cause interactions.

*J. X Wave Transformation*

Because X waves are orthogonal [61], similar to plane waves, any physically realizable waves or well-behaved solutions to the wave equation can be expressed as a linear superposition of X waves (inverse X wave transform) and the coefficients of the superposition can be determined (forward X wave transform) [46]-[47]. The inverse X wave transform is given by (Eq. (15) of [46]):

$$\Phi(\vec{r},t) = \sum_{n=-\infty}^{\infty} \int_0^{\pi/2} d\zeta \int_0^{\infty} dk\, T_{n,\zeta}(k) \Phi_{A_{n,k,\zeta}}(r,\phi,z-c_1 t)$$

$$= \sum_{n=-\infty}^{\infty} \int_0^{\pi/2} \left[ e^{in\phi} \int_0^{\infty} T_{n,\zeta}(k) J_n(kr \sin\zeta) e^{ik\cos\zeta(z-c_1 t)} dk \right] d\zeta, \quad (53)$$

$$= \sum_{n=-\infty}^{\infty} \int_0^{\pi/2} \Phi_{X_{n,\zeta}}(r,\phi,z-c_1 t) d\zeta$$

where

$$T_{n,\zeta}(k) = B_{n,\zeta}(k) e^{-ka_0}, \quad (54)$$

and

$$\Phi_{A_{n,k,\zeta}}(r,\phi,z-c_1 t) = e^{in\phi} J_n(kr \sin\zeta) e^{ik\cos\zeta(z-c_1 t)}, \quad (55)$$

where $c_1 = c/\cos\zeta$ and $|\zeta| < \pi/2$.

The forward X wave transform can be used to determine the coefficients (Eq. (26) of [46]):

$$T_{n,\zeta}(k) = \frac{k^2 c \sin\zeta \cos\zeta H(k)}{(2\pi)^2} \times \int_0^{\infty} r\,dr \int_{-\pi}^{\pi} d\phi \int_{-\infty}^{\infty} dt\, \Phi(r,\phi,z,t) \Phi^*_{A_{n,k,\zeta}}(r,\phi,z-c_1 t), \quad (56)$$

where

$$\Phi^*_{A_{n,k,\zeta}}(r,\phi,z-c_1 t) = e^{-in\phi} J_n(kr \sin\zeta) e^{-ik\cos\zeta(z-c_1 t)} \quad (57)$$

is a complex conjugate of $\Phi_{A_{n,k,\zeta}}(r,\phi,z-c_1 t)$ and $H(k)$ is the Heaviside step function [146]:

$$H(k) = \begin{cases} 1, & k \geq 0 \\ 0, & \text{Otherwise} \end{cases}. \quad (58)$$

$H(k)$ is used to indicate that $k$ is positive and thus it can be placed at either side of (56).

*K. Bowtie Limited-diffraction Beams*

If $\Phi_N(\vec{r}_N,t) = \Phi_N(\vec{r}_{N-1}, x_N - c_1 t)$ is a limited-diffraction solution to the isotropic/homogeneous wave equation (1), the Klein-Gordon Equation (20), or the Schrodinger Equation (30) (assuming that $V$ is not a function of the corresponding component of $\vec{r}_{N-1}$), where $\vec{r}_N = (x_1, x_2, \cdots x_N)$, $\vec{r}_{N-1} = (x_1, x_2, \cdots x_{N-1})$, $N$ is an integer, and $c_1$ is the speed of the wave, any partial derivatives of $\Phi_N(\vec{r}_{N-1}, x_N - c_1 t)$ along any component of $\vec{r}_{N-1}$ are still limited-diffraction solutions to these equations [147]-[151]. These solutions are called bowtie beams because their transverse beam shapes are similar to a bowtie. These beams may have applications in medical imaging of a lower sidelobe because one part of sidelobe of a transmission beam may be used to cancel the other part of sidelobe of a reception beam [147]-[151]. (Note, the following properties are also true. Any partial derivatives of a limited-diffraction solution, $\Phi_N(\vec{r}_{N-1}, x_N - c_1 t)$, in terms of the time, $t$, will also be a limited-diffraction solution to (1), (20), and (30), respectively. One example is the 2$^{nd}$ derivative X wave in terms of time given in [44]. Replacing $t$ with $-t$ in $\Phi_N(\vec{r}_{N-1}, x_N - c_1 t)$, one obtains a time reversal mirror limited-diffraction wave propagating in a backward direction along $x_N$.)

*L. Limited-diffraction Array Beams*

If the partial derivatives are carried out on more than one components of $\vec{r}_{N-1} = (x_1, x_2, \cdots x_{N-1})$ for $\Phi_N(\vec{r}_N, t) = \Phi_N(\vec{r}_{N-1}, x_N - c_1 t)$, limited-diffraction grid or layered array beams may be produce for equations (1), (20), and (30) (assuming that $V$ is not a function of the corresponding components of $\vec{r}_{N-1}$) [152]-[155]. Array beams may have applications to 3D imaging [152], blood flow velocity measurements [153], and high frame rate imaging [62]-[63], [78]-[79].

*M. Computation with Limited-diffraction Beams*

Efficient computation of limited-diffraction beams produced by a finite aperture is important for understanding the properties of these beams. A Fourier-Bessel method [24]-[28] has been used to calculate arbitrary waves of axial symmetry. Limited-diffraction array beams [152]-[155] have been used for efficient computation of waves produced by a 2D array transducer [154]-[155]. Angular spectrum decomposition has been used for the study [156] and various methods have been investigated [157].

## III. APPLICATIONS OF LIMITED-DIFFRACTION BEAMS

*A. Medical Ultrasound Imaging*

Limited-diffraction beams are localized waves and are, in theory, propagation invariant. In practice, because the dimension of wave sources is always finite, these waves will eventually diffract. However, these waves have a large depth of field, meaning that they will propagate over a large distance without spreading. This property is useful in medical ultrasound imaging where an extended depth of focus is needed to provide clear images over the entire depth of interest within the thickness of the human body. Studies on this subject have been reported, for example, in the literatures [15]-[17], [158]-[163].

*B. Tissue Characterization (Identification)*

Due to the large depth of field of limited-diffraction beams, these beams may be used for tissue characterization (identification) [164]-[166]. For example, different tissues have different attenuations on ultrasound waves. If the waves diffract as they propagate, such as conventional focused waves, one has to compensate for the diffraction effects of the waves in the estimation of tissue attenuation. The compensation process could be computationally intensive and tedious. An example of tissue characterization with limited-diffraction beams is given in [166].

*C. High Frame Rate Imaging*

High frame rate 2D and 3D ultrasound imaging is important for visualizing fast moving objects such as the heart. Based on our previous studies of ultrasound diffraction tomography [167]-[171] and limited-diffraction beams such as X waves [41]-[53], we have developed the high frame rate imaging method [62]-[83]. Recently, the method has been extended to include steered plane wave and limited-diffraction array-beam transmissions [78]-[80].

*D. Two-Way Dynamic Focusing*

A two-way dynamic focusing method was developed by transmitting limited-diffraction array beams and receiving ultrasound echo signals with array beam weightings of the same parameters. This method increases image field of view and image resolution due to enlarged coverage of spatial Fourier domain [172].

*E. Medical Blood Flow Measurements*

Blood flow velocity measurements and imaging are important for medical diagnoses [173]-[174]. However, with conventional Doppler method, only flow velocity that is along the ultrasound beam can be measured. To measure velocity vector, both the velocity components that are along with and transverse to the beam are needed. Limited-diffraction beams may help to measure the transverse component of the velocity more accurately due to their spatial modulation properties [153], [175]-[176].

*F. Nondestructive Evaluation (NDE) of Materials*

Nondestructive evaluation (NDE) is important for many applications such as finding defects in aircraft engines with ultrasound without slicing them apart or destroying them. Similar to medical imaging, limited-diffraction beams can also be applied to NDE on various industrial materials by getting images of a large depth of field [177]-[178].

*G. Optical Coherent Tomography*

Optical coherent tomography (OCT) uses the same principle of conventional ultrasound pulse-echo imaging. It is able to obtain microscopic images of a cross-section along an optical beam. Similar to ultrasound imaging, limited-diffraction beams can be applied to increase the depth of field of OCT [179].

*H. Optical Communications*

Limited-diffraction beams such as X waves [41]-[43] are orthogonal in space. Because of this property, signals such as television (TV) programs in different channels can be sent over the same space from the same channel (carrier frequency). Limited-diffraction beams have been exploited to increase the capacity in communications using the property of their spatial orthogonality [180]-[181].

*I. Reduction of Sidelobes in Medical Imaging*

Limited-diffraction beams can maintain a high resolution in medical imaging over a large depth of field. However, compared to focused beams at their focuses, limited-diffraction beams have a higher sidelobe. Sidelobes may lower image contrast in ultrasound imaging, making the differentiation between benign and malignant tissues difficult. Various methods have been developed to reduce sidelobes of limited-diffraction beam in medical imaging [5], [182]-[185].

## IV. CONCLUSION

Limited-diffraction beams are a class of waves that may be localized in both space and time and can propagate rigidly in a free space or confined spaces to an infinite distance in theory at a superluminal speed. Because of the localized property and the fact that they are solutions to various wave equations, limited-diffraction beams may provide an insight into various physical phenomena and may have theoretical significances. In addition, limited-diffraction beams can be approximately produced with a finite aperture and energy over a large depth of field, meaning that they can keep a small beam width over a large distance. This and other properties of limited-diffraction beams make them suitable for various applications such as medical imaging, tissue characterization, blood flow measurement, nondestructive evaluation of materials, and optical communications, etc.

## V. ACKNOWLEDGEMENTS

This work was supported in part by the grant, HL60301, from the National Institute of Health, USA.

## FIGURES AND LEGENGS

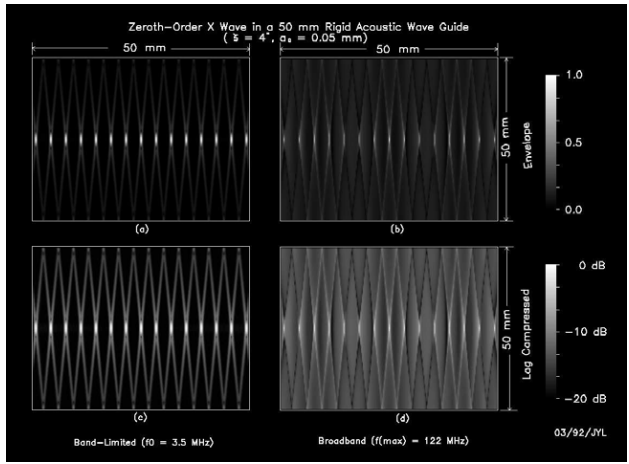

**Fig. 1.** Envelope detected Zeroth-order X Wave in a 50 mm diameter rigid acoustic waveguide. The waves shown has an Axicon angle of 4º and $a_0$ = 0.05 mm. (a) and (c) Band-limited version with a Blackman window function centered at 3.5MHz with about 81% of fractional -6dB bandwidth. (b) and (d) are a broadband version. The images in the top row are in a linear scale and those in the bottom row have a log scale to show the sidelobes.

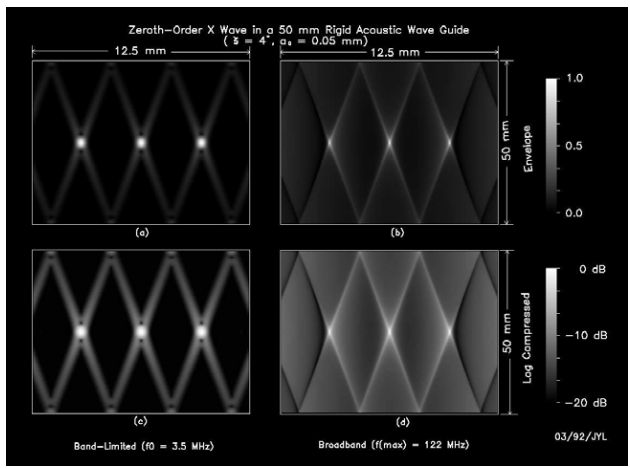

**Fig. 2.** The same as those in Fig. 1. except that the images are zoomed horizontally around the center.

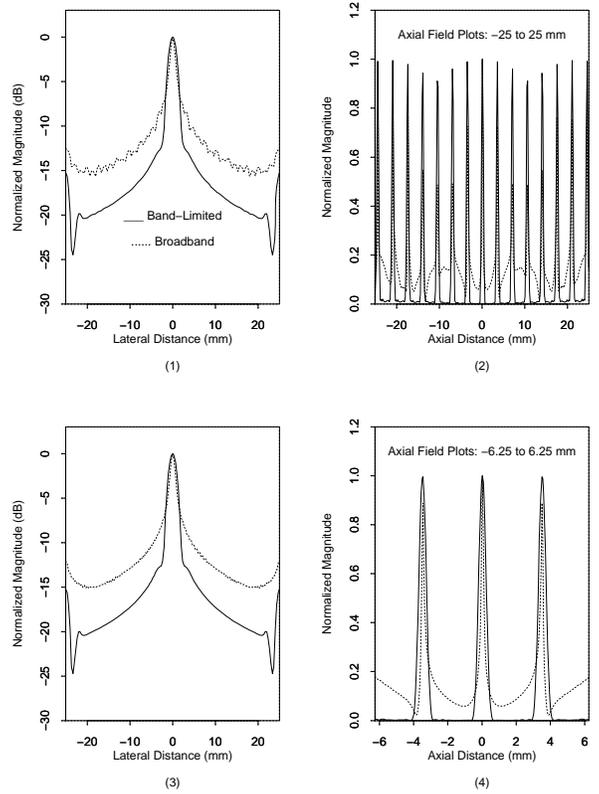

**Fig. 3.** Transverse ((1) and (3)) and axial ((2) and (4)) sidelobe plots of the images in Fig. 1. ((1) and (2)) and Fig. 2. ((3) and (4)), respectively. Solid and dotted lines are for band-limited and broadband cases, respectively.